\documentclass[conference,10pt]{IEEEtran}
\IEEEoverridecommandlockouts

\usepackage{cite}
\usepackage{makecell}
\usepackage{float}
\usepackage{amsmath,amssymb,amsfonts}
\usepackage{algorithmic}
\usepackage{graphicx}
\usepackage{textcomp}
\usepackage{xcolor}
\usepackage{newtxtext}
\usepackage{newtxmath}
\usepackage{setspace}
\usepackage{float}
\usepackage{placeins}
\usepackage{dblfloatfix}
\def\BibTeX{{\rm B\kern-.05em{\sc i\kern-.025em b}\kern-.08em
    T\kern-.1667em\lower.7ex\hbox{E}\kern-.125emX}}
\begin{document}

\title{Interpretable Stress Detection from ECG Signals Using Motif-Based Anomaly Analysis}

\author{\IEEEauthorblockN{Zhanna Balyan, Sachin Kumar}
\IEEEauthorblockA{Zaven P. and Sonia Akian College of Science \& Engineering, American University of Armenia, Yerevan, Armenia \\
zhanna\_balyan@edu.aua.am; s.kumar@aua.am}}

\maketitle

\begin{abstract}
Stress detection using physiological signals has gained significant attention due to its impact on both physical and mental health. While existing approaches based on machine learning and deep learning achieve strong predictive performance, they often rely on black-box models and fail to capture individual variability in physiological responses. In this work, we propose an interpretable and personalized framework for stress detection using electrocardiogram (ECG) signals based on motif discovery and Matrix Profile analysis. Instead of training a classifier, the method learns subject-specific baseline cardiac behavior by extracting recurring heartbeat patterns (motifs) from ECG signals. Stress is then detected as a deviation from these baseline patterns using a distance-based anomaly score.

Experiments are conducted on the WESAD dataset using a carefully designed train–validation–test protocol to ensure reliable evaluation. The results show that the proposed approach can effectively detect stress for several subjects while providing clear interpretability through direct comparison of ECG patterns. However, the performance varies across individuals due to differences in physiological responses, with some subjects exhibiting minimal morphological changes under stress. Additional analysis incorporating heart rate variability (HRV) features reveals that while HRV can improve performance in certain cases, its contribution is not consistent across all subjects.

These findings highlight the importance of personalization and interpretability in physiological stress detection and demonstrate that motif-based approaches provide a meaningful alternative to black-box models, while also revealing inherent limitations due to inter-subject variability.

\end{abstract}


\section{Introduction}
Stress is a significant factor affecting both physical and mental health, and its prolonged presence can lead to serious conditions such as cardiovascular disorders, anxiety, and reduced cognitive performance. As a result, reliable and continuous stress monitoring has become an important area of research, especially with the increasing availability of wearable sensors \cite{r1}. Among various physiological signals, electrocardiogram (ECG) signals are widely used due to their ability to capture detailed information about heart activity and their strong connection to the autonomic nervous system, which plays a key role in stress response \cite{r2}.

Existing approaches for stress detection from ECG signals largely rely on machine learning and deep learning models \cite{r1, r3}. While these methods often achieve good predictive performance, they typically operate as black-box systems, making it difficult to understand how decisions are made \cite{r4}. In sensitive applications such as healthcare, this lack of interpretability can reduce trust and limit practical adoption. In addition, many of these approaches are designed as generalized models, which may not effectively capture the natural variability in physiological signals across different individuals.

To address these limitations, this work explores an interpretable and personalized approach to stress detection using ECG signals. Instead of training a conventional classifier, we model normal cardiac behavior for each subject using motif discovery based on the Matrix Profile \cite{r5}. A motif represents the most recurring pattern in a time series and, in this context, corresponds to a typical heartbeat pattern during a relaxed state \cite{r6}. Stress is then identified as a deviation from this learned baseline by measuring the dissimilarity between the incoming ECG segments and the extracted motifs.

The proposed approach offers three main advantages. First, it provides interpretability by allowing for direct comparison between normal and anomalous heartbeat patterns. Second, it supports personalization by learning subject-specific baselines rather than relying on a single global model. Finally, it operates entirely unsupervised, which removes the need for pre-labeled stress data for identifying physiological changes.
Through experiments on ECG data, we demonstrate that stress-related segments exhibit a higher deviation from baseline motifs, highlighting the effectiveness of the approach.

\section{Literature Review}

Stress is a predominant contributing factor to mental and physical deterioration. Often, unnoticed and subtle, it can lead to acute conditions such as cardiovascular disease, depression, brain atrophy, and cognitive impairment\cite{r8}. Stress activates the autonomic nervous system and causes changes in heart rate, rhythm, and the morphological shape of the heartbeat.
Stress detection using physiological signals has received significant attention due to its importance in healthcare and well-being. Among various signals, ECG is widely used because it captures detailed information about cardiac activity and reflects the response of the autonomic nervous system. Many studies have used features derived from ECG, such as heart rate and heart rate variability, to distinguish between relaxed and stressed states \cite{r1,r2}.
The most widely studied approach to stress detection using ECG signal data was through the application of machine learning and deep learning models. Methods based on classifiers such as support vector machines, random forests, and neural networks have shown promising results \cite{r3,r5}.
Particularly, in the case of binary stress classification, Schmidt et al. achieved an accuracy of 0.85 and an F1-score of 0.81 for their baseline evaluation applying Linear Discriminant Analysis (LDA) only on ECG data. An F1-score of 0.91 was achieved only after the inclusion of all chest-worn physiological metrics \cite{r3}.

More recent studies have relied on deep learning models that transform ECG signals into 2D images. For example, Sriram Kumar et al. used a pre-trained VGG16 network and achieved an F1-score of 0.74 on the WESAD dataset \cite{r9}. Yet, the performance of the model was highly sensitive to its architecture, as other approaches like AlexNet or configurable CNNs reached substantially lower F1-scores, like 0.48 or 0.55. 
However, these approaches often operate as black-box models, making it difficult to interpret how decisions are made \cite{r4}. This lack of transparency is a limitation, especially in applications where understanding the reasoning behind predictions is important.

Another limitation of existing approaches is that they are often designed as generalized models. Physiological signals, including ECG, can vary significantly across individuals due to differences in body composition, cardiovascular health or previous exposure to stress. For this reason, a single global model may not effectively capture the differences in ECG signals \cite{r5}. 
This highlights the need for approaches that are not only accurate but also personalized.
The importance of personalization was revealed in another study conducted on the WESAD dataset, which applied CNN-based approaches using both a generalized approach and a personalized approach. The difference between the two approaches was drastic because the personalized approach reached an F1-score of 0.91, while the generalized approach reached an F1-score of only 0.43 \cite{r10}.

An alternative way to approach this problem is to model normal physiological behavior and detect deviations from it. This shifts the problem from classification to anomaly detection, where stress can be viewed as a deviation from a subject’s baseline cardiac activity. Such an approach is naturally more aligned with personalized modeling.

In time series analysis, motif discovery has been widely used to identify recurring patterns, while the Matrix Profile has emerged as an efficient tool for measuring similarity and detecting anomalies \cite{r6,r7}. 
These methods allow for direct comparison between patterns in a signal and provide a clear interpretation of similarity and deviation.

However, the use of motif-based and matrix profile techniques for interpretable and personalized stress detection from ECG signals remains relatively unexplored. While there is plenty of research on how traditional supervised approaches achieve high accuracy on average, they have a crucial limitation of being non-transparent, dependent on labels, and generalized. On the other hand, there is limited research about applying the Matrix Profile approach for detecting anomalies in ECG signals. Wankhedkar and Jain explored this method and established that changes in heartbeat morphology can be detected with a distance-based approach, instead of applying heavy manual feature engineering \cite{r11} .  
This motivates the present work, where we aim to develop a method that builds upon the previously mentioned research and combines interpretability, personalization, and effective stress detection.

\section{Methodology}

\subsection{ECG Signal Description}

The Electrocardiogram (ECG) is a time-voltage chart that represents the electrical activity of the heart from one instant to the next during each contraction and relaxation cycle \cite{r12}.

\begin{figure}[htbp]
    \centering
    \includegraphics[width=0.4\textwidth]{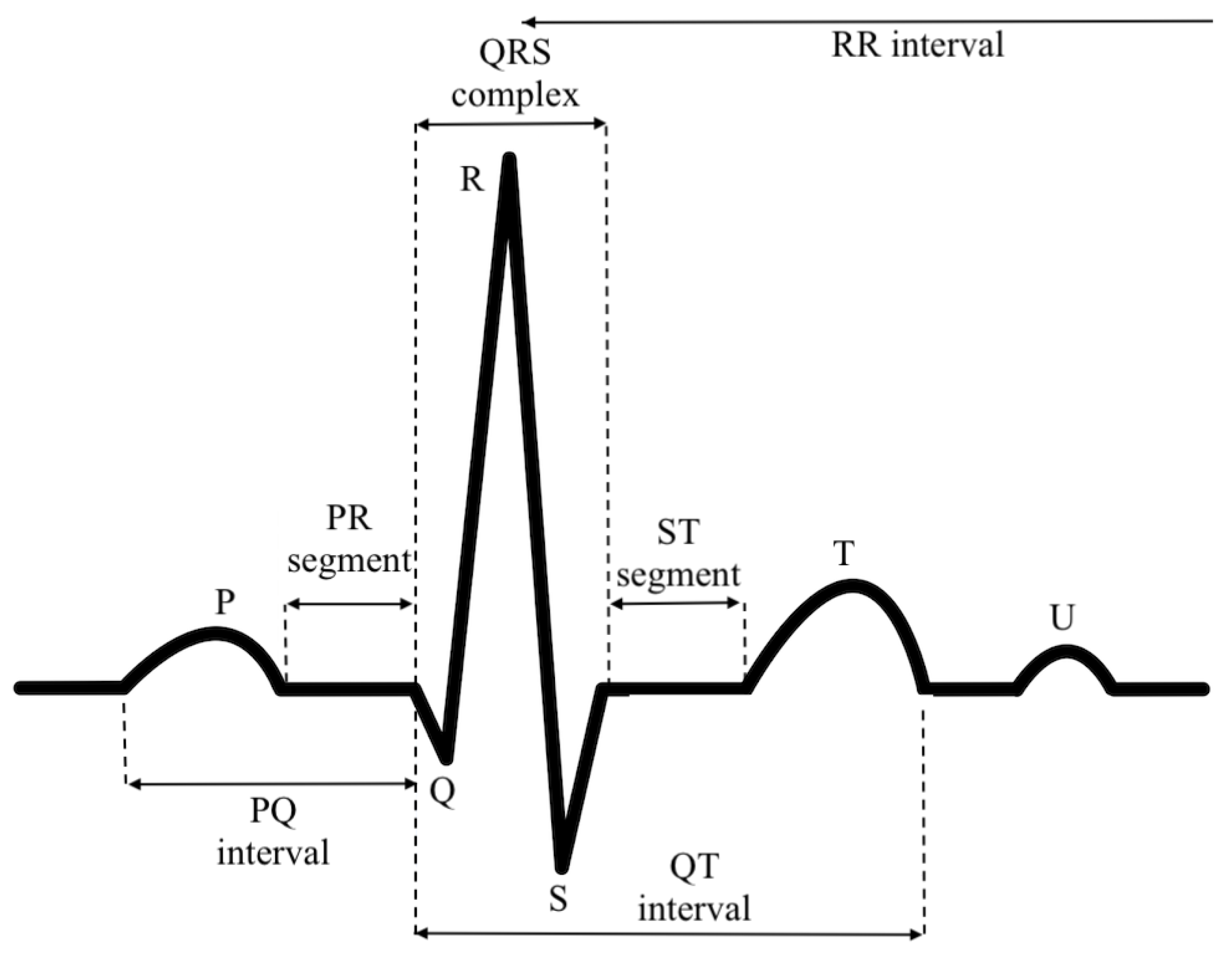}
    \caption{Typical morphological components of the ECG waveform.}
    \label{fig:ecg_segment}
\end{figure}

As shown in Figure \ref{fig:ecg_segment}, every heartbeat produces a typical waveform that consists of three main morphological components. The P wave describes the initial deviation, which is the process of the contraction of the heart’s upper chambers, otherwise known as atrial depolarization. The QRS complex is responsible for producing the most sharply defined feature of the ECG with the help of the contraction of the lower chambers of the heart, a process known as ventricular depolarization. The final wave called the T wave, represents the phase during which the heart muscle resets for the next beat, which is called ventricular repolarization.  

Under stable baseline conditions, the P-QRS-T waveform repeats with recurring consistency across consecutive cardiac cycles. However, when a person experiences stress, their autonomic nervous system induces changes in the morphological shape of the heartbeat, including height changes in the QRS complex or sudden shifts in the T-wave. Consequently, as the heart’s shape no longer aligns with its relaxed baseline, these divergences mark the changes as anomalies and facilitate our model to detect stress with the help of an unsupervised approach.

\subsection{Motifs}

The concept of time-series motifs was first introduced to catch previously unknown recurring patterns in sequential data \cite{r14}. It is formally defined as a pair of subsequences ($T_{i,m}, T_{j,m}$) that are the most identical to each other across all possible pairs of non-overlapping subsequences of the same length $m$, given a specific distance measure \cite{r15}.

This concept was further expanded to extract a top-$K$ motif set, which represents $K$ most repeating structural patterns that are associated with a specific motif. In contrast, subsequences that are the furthest from one another and have no aligned matches within the signal are called time-series discords or anomalies \cite{r16}.

By looking at above-mentioned time-series concepts from a physiological perspective, it becomes evident that motifs naturally correspond to the frequently occurring cardiac P-QRS-T waveforms. Furthermore, a set of motifs can accurately detect the most persistent and dominant shapes and patterns appearing in the subject’s baseline ECG, which makes motif discovery an effective and interpretable approach in order to learn what a relaxed cardiac activity looks like for every person. In the same way, stress-induced morphological changes in the patterns of the heartbeat can be seen as subsequences or discords deviating from these learned patterns, which frames the process as an anomaly detection.

\subsection{Z-Normalized Euclidean Distance}

In order to define whether two subsequences are considered motifs or anomalies, this framework utilizes the Z-Normalized Euclidean distance between two subsequences of length $m$ as its core distance measurement metric. Before comparison, each subsequence is first standardized by subtracting its mean and scaling by its standard deviation, resulting in a uniform representation with zero mean and unit variance. 

This is a crucial preprocessing step in detecting morphological changes, as raw ECG recordings are extremely sensitive to amplitude fluctuations and baseline wander, which can be brought about by natural physiological events like breathing, sweating, or other such movements. When z-normalization is applied, these fluctuations are discarded, making sure the framework doesn’t classify two similar heartbeats as divergent and evaluates the true shape of each cardiac waveform. 

As the next step, the z-normalization is followed by the calculation of the distance between two subsequences of length $m$ with the help of the following equation: 

\begin{equation}
    d_{i,j} = \sqrt{2m \left( 1 - \frac{T_{i,m} \cdot T_{j,m} - m\mu_i\mu_j}{m\sigma_i\sigma_j} \right)}
\end{equation}

where:

\begin{itemize}
    \item $d_{i,j}$ is the distance score between the two subsequences.
    \item $m$ is the size of the temporal window or the length of the subsequences.
    \item $T_{i,m}$ and $T_{j,m}$ are the subsequences that are being compared.
    \item $T_{i,m} \cdot T_{j,m}$ represents the dot product of the two subsequences.
    \item $\mu_i, \mu_j$ are the mean values of the respective subsequences.
    \item $\sigma_i, \sigma_j$ are the standard deviations of the respective subsequences.
\end{itemize}

\subsection{Matrix Profile}
With a robust distance metric established for morphological comparison, the framework requires an effective computational mechanism to evaluate the structural resemblance of all possible subsequences across the entire time series. To execute this global similarity search, the system first calculates distance profiles and subsequently constructs a Matrix profile \cite{r7}. 

A Distance Profile ($D_i$) is a one-dimensional array that stores the z-normalized Euclidean distance between a specific time series subsequence $T_{i,m}$ and every other subsequence in the entire time series $T$. 
Mathematically, the Distance Profile is defined as:
\begin{equation}
    D_i = [d_{i,1}, d_{i,2}, \dots, d_{i,n-m+1}]
\end{equation}
where $n$ and $m$ represent the length of the entire time series and the length of the specific subsequence, respectively, and $d_{i,j}$ (for $1 \le i, j \le n-m+1$)   represents the distance between the subsequence $T_{i,m}$ and the target subsequence $T_{j,m}$. \cite{r7}. 

In order to apply this concept to the entire signal, the Matrix Profile \cite{r7} is introduced, which is a data structure that encodes the z-normalized Euclidean distance between each subsequence $T_{i,m}$ and its closest match in the entire signal. Matrix Profile is represented as:
\begin{equation}
    P = [\min(D_1), \min(D_2), \dots, \min(D_{n-m+1})]
\end{equation}
where every term $\min(D_i)$ (for $1 \le i \le n-m+1$) defines the absolute minimum value taken from its respective distance profile $D_i$. When these minima are extracted, the matrix profile guarantees an efficient similarity comparison by discarding irrelevant values and maintaining only the closest matches. 

Additionally, the Matrix Profile Index ($I$), which is a vector of integers, stores the temporal index that points to the exact location of the nearest neighbor occurring within the recording. 
By mapping these concepts to our ECG signal, the matrix profile reveals the recurring motifs through its lowest values (global minima) and discords through the highest peak values (maxima).

\subsection{Dataset Description}
In the scope of the proposed framework, the Wearable Stress and Affect Detection (WESAD) \cite{r3} dataset was utilized, which is a multimodal dataset that incorporates physiological and motion data recorded with the help of a chest-worn device (RespiBAN) and a wrist-worn device (Empatica E4). The data was retrieved from 15 healthy subjects, all of whom underwent a rigorously monitored laboratory procedure which was intentionally designed to capture a stable and clear physiological baseline followed by subsequent responses to acute stress stimuli.  
While WESAD included a vast variety of different physiological signals, our research puts an emphasis on the high-resolution Electrocardiogram (ECG) data collected from the chest-worn RespiBAN device. The ECG signals were taken at a sampling rate of 700 Hz to ensure that enough details were captured in order to have a clear picture of the heartbeat for more accurate detection of morphological changes.

\subsection{Experimental Protocol}

The subjects of the experiment were given multiple instructions prior to the experiment such as avoiding caffeine, tobacco, and excessive strenuous exercise before the study. Additionally, they were carefully guided through every step of the study in order to induce three affective states, including neutral, amusement, and stress.

During the first phase of the study, which aimed to induce a relaxed affective state, the subjects were given neutral reading materials and were recorded sitting or standing beside a table. In the scope of our proposed framework, the ECG data taken during this phase of the experiment serves as a foundation for obtaining the recurring baseline patterns or motifs, which characterize the subject’s standard cardiac behavior before the introduction of stress stimuli.

Following the neutral state, the subjects went through the second phase of the protocol, which was the amusement state. This phase was induced by having the subjects watch a set of eleven funny clips. However, since the foremost objective of our framework was binary stress detection, the data collected during this phase of the experiment wasn’t utilized.

Instead, the study we propose focuses completely on the third phase, during which the subjects experienced acute psychological stress. To accurately induce this condition, the subjects were exposed to the Trier Social Stress Test (TSST), which included delivering an impromptu speech to a panel of judges followed by an arithmetic task of having to serially subtract 17 from 2023 \cite{r3}. The ECG segments recorded during this phase of the experiment act as the primary test data for anomaly detection. 

\subsection{ECG Signal Preprocessing}

As an initial step, the ECG data were subjected to a standardized preprocessing pipeline before motif discovery in order to ensure the integrity of the process. The preprocessing pipeline included two crucial steps, which were aimed at removing external noise, such as physiological artifacts that originated from other biological processes of the body and could potentially harm the process of pattern analysis and recognition. 

A third-order Butterworth bandpass filter with a passband of 0.5 - 45 Hz was applied in order to isolate and preserve only the relevant components of the cardiac cycle. The particular passband was chosen in alignment with clinical monitoring standards \cite{r17}. The lower cutoff frequency of 0.5 Hz was chosen to eliminate baseline wander, which includes low-frequency drift that can be caused by the subject’s respiration and can cause artificial fluctuation. Meanwhile, the higher cutoff frequency of 45 Hz was chosen in order to remove high-frequency noise like muscle-induced electromyographic signals. The filter was applied bidirectionally to prevent phase distortion of the P, QRS, and T wave components.

As the next step, the filtered signal was subjected to Z-score normalization by subtracting the mean and dividing by its standard deviation in order to transform the ECG signal into a dimensionless signal with zero mean and unit variance. This step is performed to get rid of non-cardiac factors that can cause inconsistencies in the signal. Instead, we make sure the framework focuses solely on the true morphological shape of the cardiac waveform.

\subsection{Data Separation}
Following the data preparation step, each subject’s preprocessed ECG was separated into three strategic splits for the purpose of preventing data leakage. 
The first split was allocated for the training phase, which served as the basis for motif learning and was utilized exclusively for that purpose. This partition consisted of the first five minutes of baseline ECG data corresponding to 210.000 samples recorded at 700 Hz.

The next three minutes of baseline data, along with the first three minutes of the stress data consisting of 126.000 samples each, were merged together into the validation set. This data segment was used entirely for threshold calibration with the help of a systematic percentile search process that will be described in the later sections. By incorporating equal lengths of baseline and stress data, we made sure the framework didn’t get prematurely exposed to the data dedicated for evaluation and picked a fair cutoff point.

Finally, the last split of the data included the remaining samples of both baseline and stress conditions and served for the computation of the final evaluation metrics. For example, in the case of the second subject, this data consisted of approximately 11 minutes of baseline and 7 minutes of stress data samples. 

By partitioning the data into three distinct segments, we help the framework learn motifs before threshold calibration and define a threshold before the test split evaluation, which ensures that no data samples from a later phase flow back into an earlier one.

\subsection{Matrix Profile Computation}

The computations entailed in the Matrix Profile were executed with the help of STUMPY, which is an effectively optimized Python library developed for data mining tasks \cite{r18}. This library was chosen because it is highly efficient in calculating z-normalized Euclidean distances within large volumes of data.
The most significant parameter that makes the Matrix Profile computation possible is the temporal window size chosen for all subjects and denoted by m. After careful experimentation, the length of the window size was chosen to include 1050 samples, which corresponds to 1.5 seconds of ECG data recorded at a rate of 700 Hz. This selection wasn’t arbitrary because, from a physiological standpoint, a window size of 1.5 seconds allows one to capture exactly one distinct heartbeat. If the resting heart rate of one is 60 to 80 beats per minute, it signifies that the full cardiac cycle lasts from 0.8 to 1.2 seconds, and the selected window size allows us to capture the morphological structure of one P-QRS-T waveform. Moreover, the size of the window is adequate to make sure that two consecutive beats aren’t included during a single window.

\subsection{Motif Extraction}

The first split of the data was utilized for gaining insights about the baseline of normal cardiac behavior. From the Matrix Profile calculation described above, the top five distinct motifs were chosen by applying a greedy exclusion zone method. The values computed in the Matrix Profile were sorted from lowest to highest in order to make it possible to detect recurrent subsequences, and the one with the lowest score was chosen as the principal motif. Additionally, the greedy exclusion zone prevented the framework from choosing adjacent or redundant subsequences to make the detection process informative. The process was then repeated until the five motifs were also chosen. 
Since the chosen motifs were learned entirely from each subject’s personal baseline data, the framework becomes personalized as it captures each person’s specific P-QRS-T waveform shape instead of using a general approach.

\subsection{Distance Profile Computation and Smoothing}

After successfully extracting the five baseline motifs, the unseen ECG segments were evaluated through the calculation of an anomaly score by finding the minimum z-normalized Euclidean distance of a given new ECG segment to any of the chosen baseline motifs. This was done by means of Mueen’s Algorithm for Similarity Search (MASS) \cite{r18}, which is a function included in the STUMPY library and is specifically calibrated to effectively compare a pattern across large volumes of data. A separate distance profile was generated for each of the learned motifs, after which the lowest score across each time step was collected in order to show the signal’s distance to the closest motif at any moment. 

Following this step, all the distance scores were divided by the window size to make sure they can be compared with different window configurations without getting artificially inflated in case of a longer sequence length. As Euclidean distances are essentially cumulative sums of pairwise differences, the distance scores naturally scale with the size of the window, and the anomaly score can be inflated without this normalization step. In this way, we convert the score into an average distance per sample and reduce mathematical bias during the threshold calibration phase. 

Furthermore, the normalized distance profile underwent a smoothing step in order to remove transient noise and stabilize the scores. Since the state of stress lasts for several seconds or minutes, a 3-second rolling average smoothing function was applied to the distance profile to remove any brief, momentary spikes and only detect genuine stress responses.

\subsection{Threshold Calibration}

When the distance profile was acquired, the next step was to select a decision boundary that could efficiently differentiate between baseline and stress segments of the recording. Considering that our framework aims to be personalized, instead of defining a fixed value, we implemented a dynamic approach that would find the optimal threshold for each subject. 
This approach incorporated a grid search, which searched for a threshold over a large set of percentiles ranging from the 50th to the 99th percentile of the baseline distribution of the validation set. After this, the threshold with the highest general F1-score was selected as the most optimal one for the given subject. This approach allowed the system to adjust to each person’s unique cardiac behavior ensuring that even small morphological changes could have the chance to get caught. 

\subsection{Implementation Environment}

The proposed system was developed and implemented entirely in Python in order to have the ability to have computational efficiency and a reproducible workflow. Data processing and manipulation were managed by NumPy, Pandas and SciPy libraries. In order to visually detect morphological changes in the ECG signal, Seaborn and Matplotlib libraries were used to generate graphs and visuals. Finally, the system was evaluated with the help of Scikit-learn, and the Matrix profile was calculated using the STUMPY library. The full list of utilized libraries and versions is provided in the given technical documentation.
All the experiments carried out in the scope of our system were conducted on a local CPU-based machine. This choice was made due to the fact that the STUMPY library is already constructed in a way that can support parallel processing across many cores and does not require GPU acceleration.

\section{Results}

\subsection{Quantitative Results}

To ensure the framework’s efficiency in accurately identifying repeating patterns in a person’s cardiac behavior as stable, healthy morphology and consequently flagging shapes that differ as anomalies that indicate stress, it was vital to correctly choose three significant hyperparameters. Those were the window size, the number of motifs, and an optimal threshold.
First, it was important to establish a proper window size as it plays an essential role in ECG signal processing and must be chosen in a way that balances the trade-off between frequency resolution and time resolution \cite{r19}.  
Finding the optimal window size remains to this day one of the most challenging tasks in unsupervised time-series analysis, and there is no universal domain-agnostic approach that can be applied to choose an optimal window size when working with ECG signal processing  \cite{r20}. 
While some research advocates that choosing a larger window size may improve anomaly detection by capturing longer temporal structure and rhythm information \cite{r21,r22}, it can also make the system sensitive to short-term variations. Other studies state that for average heart rates, a sliding window of more than three seconds is not a beneficial choice, since it will contain more than 3-5 beats \cite{r23}. 
Taking into account the limited research mentioned above, we conducted extensive preliminary experiments during the initial development stage, motivated by the fact that the choice of window size might benefit from subject-specific tuning, as it is dependent on individual characteristics above all else. This included constructing a grid search over all three hyperparameters, which included $m \in \{350, 700, 1050\}$, motif counts of $k \in \{1, 3, 5\}$, and threshold percentiles of 90, 95, 98.  
Across all the subjects, the highest performing combination of parameters on average was a window size of 1050 (1.5 seconds) and a motif count of 5. This further reinforced the idea that the window size should be large enough to capture at least one full cardiac cycle, which can’t always be achieved with window sizes of 350 and 700. Therefore, these parameters were fixed and applied in the implementation of the final pipeline. 
However, we decided to take another approach with threshold selection because the distance values differed substantially across subjects, which made the selection of one global threshold unreliable. Instead, thresholds were calibrated individually per subject as prior research indicates that personalized monitoring improves sensitivity to subject-specific characteristics\cite{r24}.
This was done with the help of another grid search, which spanned across values between the 50th and 99th percentiles of the baseline distance distribution within the validation set. For each of the percentile values, their corresponding F1-scores were computed, which helped the cutoff selection maximize detection performance on an individual level. 
With this approach, the model remained sensitive to subtle morphological changes while also maintaining baseline ECG segments currently classified. If a fixed global threshold had been chosen, the model would have struggled to keep this balance across all the subjects. Furthermore, this would've led to missed anomalies or false positives.
The finalized hyperparameter configurations derived from the established grid search are summarized in Table \ref{tab:parameter_summary}. 

\begin{table}[htbp]
\centering
\footnotesize
\caption{Finalized Hyperparameter Configurations per Subject}
\label{tab:parameter_summary}
\begin{tabular}{lccc}
\hline
\textbf{Subject} & \textbf{$m$} & \textbf{$k$} & \textbf{Threshold (\%ile)} \\
\hline
S2  & 1050 & 5 & 93 \\
S3  & 1050 & 5 & 92 \\
S4  & 1050 & 5 & 59 \\
S5  & 1050 & 5 & 56 \\
S6  & 1050 & 5 & 71 \\
S7  & 1050 & 5 & 50 \\
S8  & 1050 & 5 & 50 \\
S9  & 1050 & 5 & 50 \\
S10 & 1050 & 5 & 91 \\
S11 & 1050 & 5 & 50 \\
S13 & 1050 & 5 & 50 \\
S14 & 1050 & 5 & 86 \\
S15 & 1050 & 5 & 50 \\
S16 & 1050 & 5 & 88 \\
S17 & 1050 & 5 & 77 \\
\hline
\end{tabular}
\end{table}

After the selection of personalized parameters, the framework’s efficiency and robustness were evaluated on the unseen test data, the results of which can be seen in the results of which can be seen in Table~\ref{tab:performance_results}. 

\begin{table}[t]
\centering
\footnotesize
\caption{Test Set Performance per Subject}
\label{tab:performance_results}
\begin{tabular}{lccccr}
\hline
\textbf{Subj.} & \textbf{Acc} & \textbf{Prec} & \textbf{Rec} & \textbf{F1} & \textbf{Sep.} \\
\hline
S2  & 0.885 & 0.921 & 0.842 & 0.879 & 1.03x \\
S3  & 0.870 & 0.921 & 0.809 & 0.862 & 1.02x \\
S4  & 0.726 & 0.689 & 0.823 & 0.750 & 1.00x \\
S5  & 0.747 & 0.676 & 0.947 & 0.789 & 1.00x \\
S6  & 0.700 & 0.641 & 0.909 & 0.752 & 1.00x \\
S7  & 0.483 & 0.483 & 0.462 & 0.472 & 1.00x \\
S8  & 0.254 & 0.065 & 0.037 & 0.047 & 0.99x \\
S9  & 0.503 & 0.505 & 0.343 & 0.408 & 1.00x \\
S10 & 0.841 & 0.768 & 0.976 & 0.860 & 1.02x \\
S11 & 0.548 & 0.537 & 0.700 & 0.608 & 1.00x \\
S13 & 0.493 & 0.486 & 0.248 & 0.329 & 1.00x \\
S14 & 0.862 & 0.855 & 0.871 & 0.863 & 1.01x \\
S15 & 0.555 & 0.552 & 0.587 & 0.569 & 1.00x \\
S16 & 0.896 & 0.927 & 0.859 & 0.892 & 1.02x \\
S17 & 0.658 & 0.700 & 0.553 & 0.618 & 1.01x \\
\hline
\textbf{Avg} & \textbf{0.668} & \textbf{0.648} & \textbf{0.664} & \textbf{0.646} & \textbf{1.01x} \\
\hline
\end{tabular}
\end{table}

The performance of the framework was assessed with the help of key metrics like Accuracy (the proportion of correctly classified physiological states), Precision (the model’s ability to avoid false positives), Recall (the model’s ability to catch true stress events), F1-score (the balance between precision and recall) and Separation ratio (the ratio of the mean anomaly score during stress to the mean anomaly score during baseline). The latter serves as a personalized score indicating how clearly the model can differentiate between baseline and stress states.

\subsection{Qualitative Results}

While the quantitative metrics comprehensively evaluate the framework’s performance, a qualitative visual analysis was also conducted to better illustrate how the Matrix Profile differentiates between the two states. By comparing the performance of a high responder like Subject 2 with the performance of a non-responder such as Subject 8, we can clearly identify what caused the framework to succeed in some cases and fail in others.

At first, we examine the baseline motifs extracted from the relaxed cardiac signal of the subjects, which can be seen in Fig.~\ref{fig:motifs_comparison}.

\begin{figure}[htbp]
    \centering
    \includegraphics[width=\columnwidth]{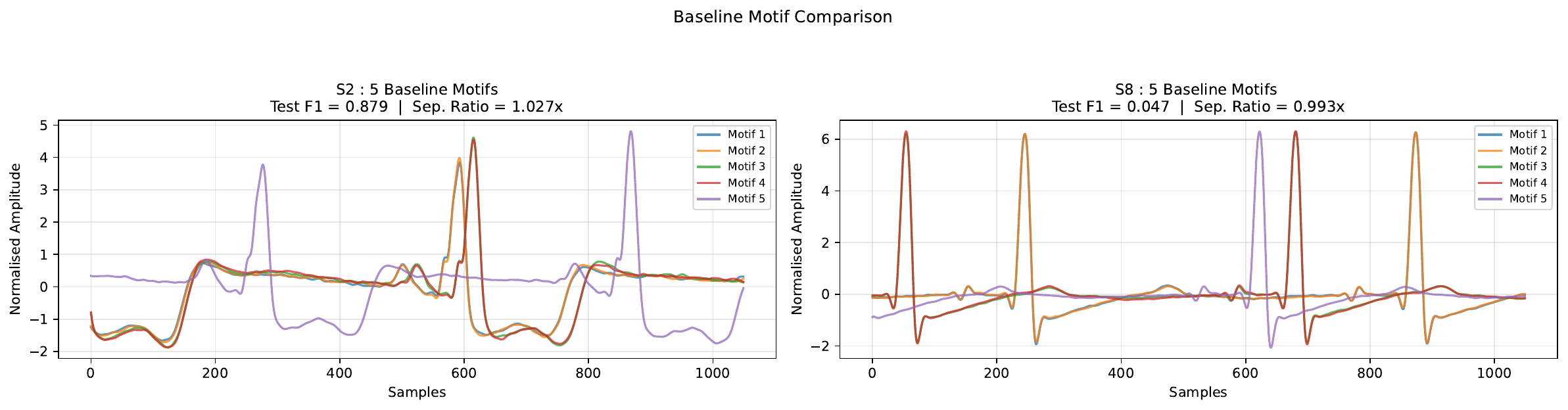}
    \caption{Baseline motif comparison between S2 (best performer, F1~=~0.879) and S8 (worst performer, F1~=~0.047).}
    \label{fig:motifs_comparison}
\end{figure}

Following this, the second step of obtaining the qualitative results entails the examination of the exact moment the anomaly detection was maximized.

\begin{figure}[htbp]
    \centering
    \includegraphics[width=\columnwidth]{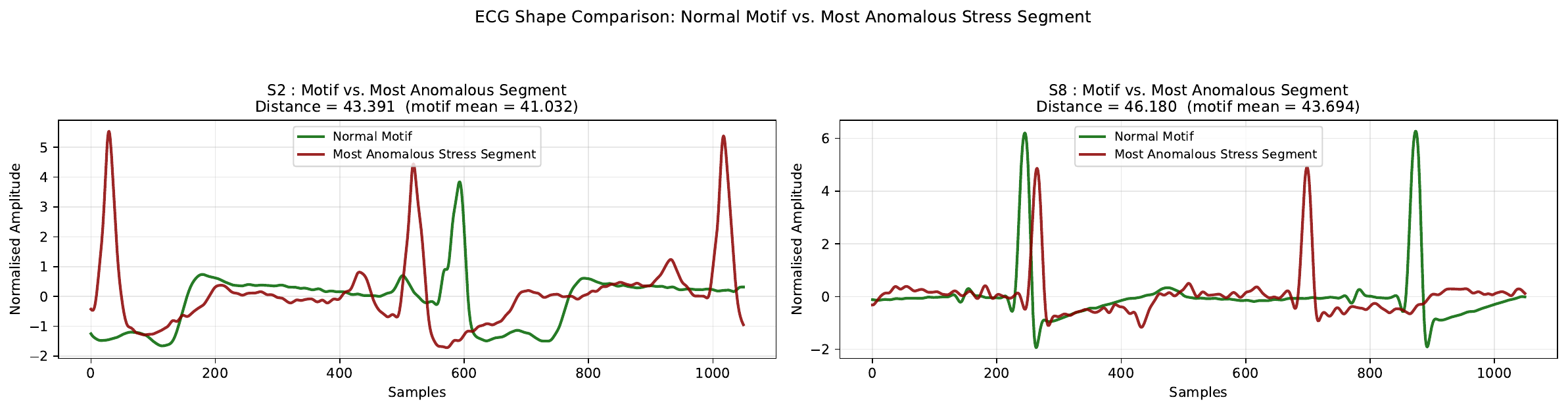}
    \caption{Baseline Motif vs Most Anomalous Segment Comparison between S2 (best performer, F1~=~0.879) and S8 (worst performer, F1~=~0.047).}
    \label{fig:shapes_comparison}
\end{figure}

Finally, the last stage of qualitative evaluation included the real-time anomaly detection analysis for the two subjects.

\begin{figure}[htbp]
    \centering
    \includegraphics[width=\columnwidth]{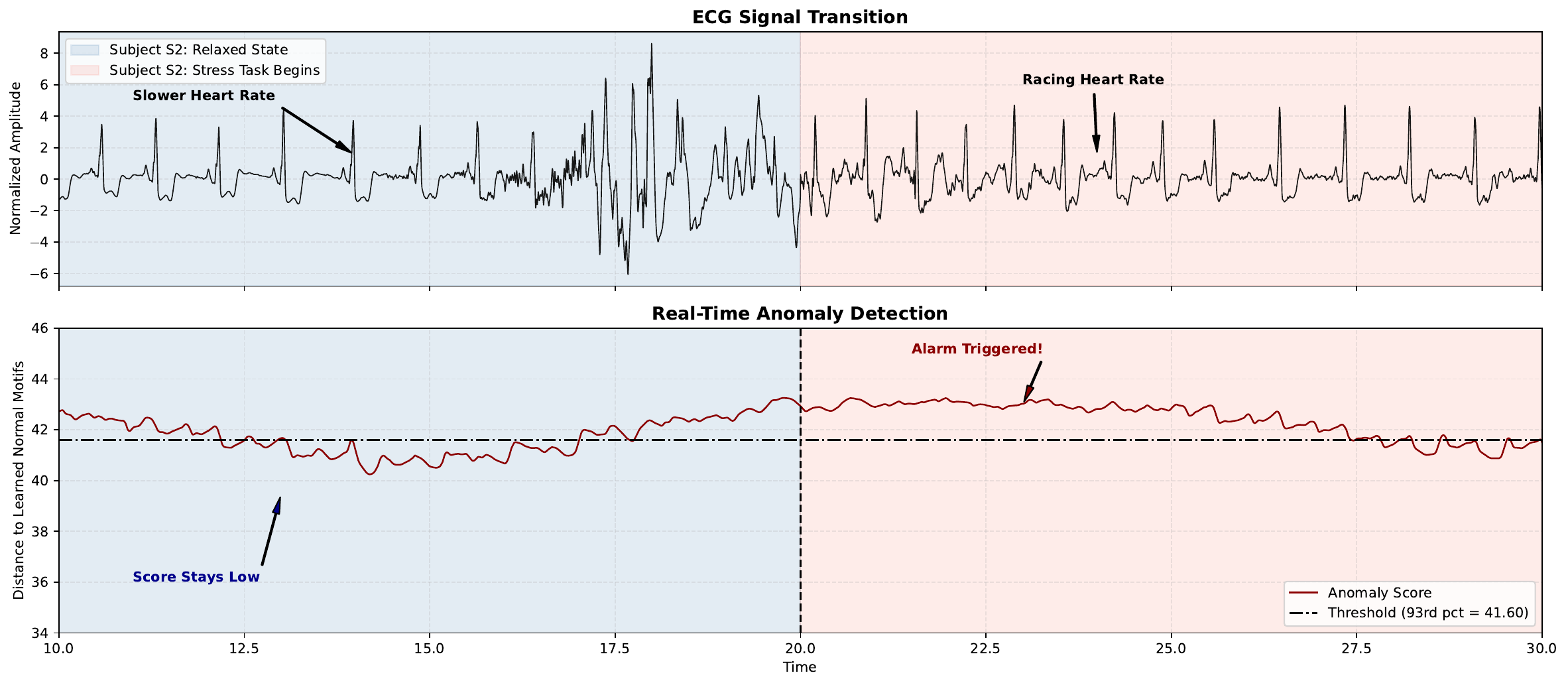}
    
    \vspace{0.1in} 
    
    \includegraphics[width=\columnwidth]{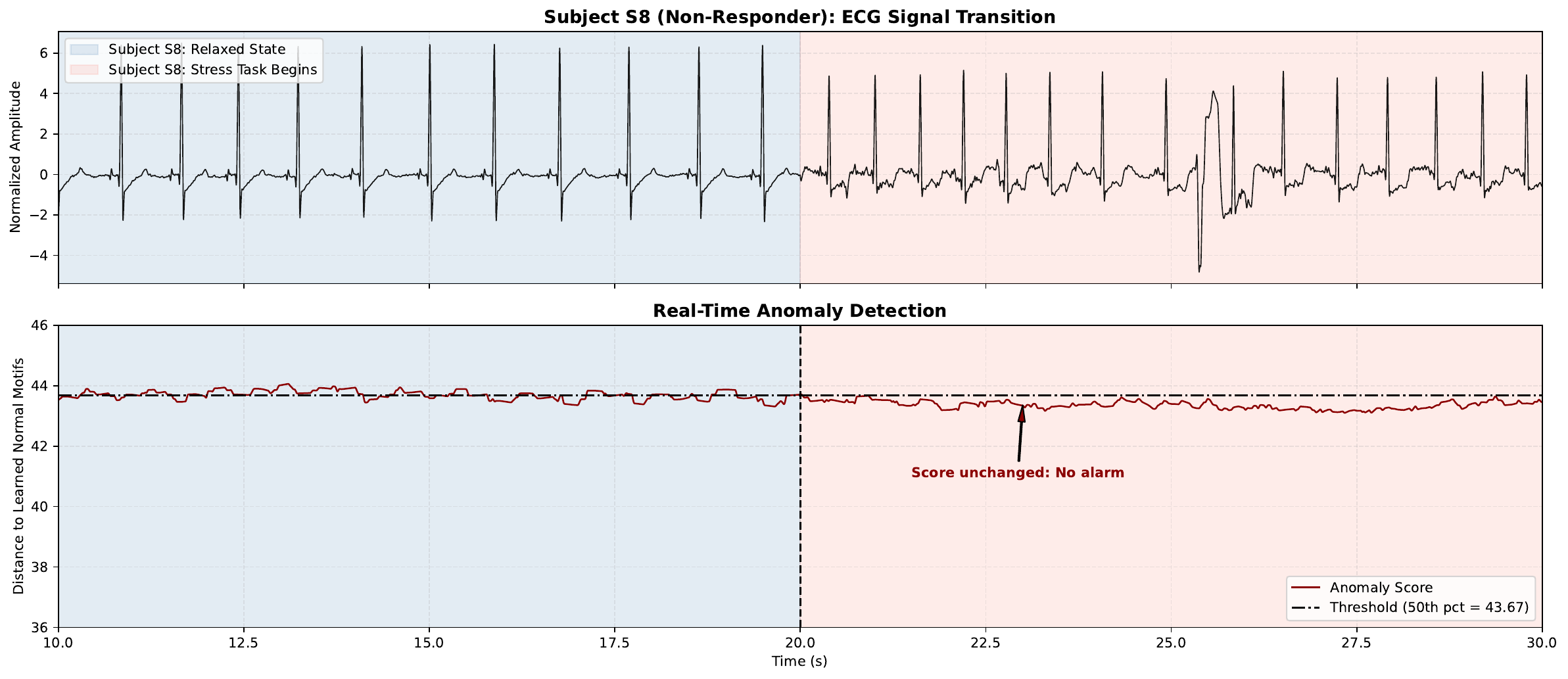}
    
    \caption{Real-time anomaly detection comparison between S2 (top) and S8 (bottom).}
    \label{fig:anomaly_detection_comparison}
\end{figure}

\subsection{Supplementary Experiment Results}

After concluding the results of the stress detection pipeline using the primary approach, we decided to conduct a supplementary experiment to understand if the addition of Heart Rate Variability (HRV) features could improve the limitations connected with the motif-based approach. More specifically, the main objective of the supplementary experiment was to observe if the performance of the subjects, where ECG morphology alone wasn’t sufficient to distinguish between the two states, would improve. 
Heart Rate Variability (HRV) consists of changes in time intervals between consecutive heartbeats that are also known as RR intervals \cite{r25}. Under stressful conditions, a person’s heart starts beating faster, and their cardiac rhythm becomes more irregular.
Rather than using multiple HRV features, the experiment was conducted using only the Root Mean Square of Successive RR differences (RMSSD). RMSSD calculates the tiny shifts in time from one heartbeat to the next. It is the most efficient metric used by researchers to measure the activity of the parasympathetic nervous system \cite{r25}. Other HRV features are all calculated from the same common heartbeat sequences, which means that combining all of them together would result in data redundancy and inflation of accuracy. For this reason, RMSSD was separated from the other features and used independently.

This supplementary experiment utilized a 15-second sliding window with a 5-second step. So, the model looked at the first 15 seconds of the data, calculated the RMSSD and the Motif distance, and then went forward by 5 seconds to look at the next portion of the data. 
Since these two features are substantially different from one another, they couldn't have been directly combined as the Motif Distance would override the RMSSD. To solve this issue, a Standard Scaler was applied to bring the two features to the same scale. Another important nuance is that when a person enters a stressed state, their RMSSD lowers, while the opposite happens with the Motif Distance. Therefore, we flipped the RMSSD, making it negative. After this, the two values were averaged together with the same weight to calculate the anomaly score. Aside from these unique methodological characteristics, the other steps remained the same as in the main approach. 

Table~\ref{tab:hrv_comparison} summarizes the final results of the supplementary experiment. 

\begin{table}[t]
\centering
\footnotesize
\caption{Performance Comparison: Motif-Only vs. Motif + HRV Features}
\label{tab:hrv_comparison}
\begin{tabular}{lccc}
\hline
\textbf{Subject} & \textbf{F1 (Motif)} & \textbf{F1 (Motif + HRV)} & \textbf{Difference} \\
\hline
S2  & 0.879 & 0.932 & +0.052 \\
S3  & 0.861 & 0.989 & +0.127 \\
S4  & 0.750 & 0.941 & +0.191 \\
S5  & 0.789 & 0.915 & +0.126 \\
S6  & 0.752 & 0.663 & -0.088 \\
S7  & 0.472 & 0.236 & -0.236 \\
S8  & 0.047 & 0.000 & -0.047 \\
S9  & 0.408 & 0.189 & -0.219 \\
S10 & 0.860 & 0.724 & -0.136 \\
S11 & 0.608 & 0.930 & +0.322 \\
S13 & 0.329 & 0.618 & +0.289 \\
S14 & 0.863 & 0.979 & +0.116 \\
S15 & 0.569 & 0.361 & -0.208 \\
S16 & 0.892 & 0.921 & +0.029 \\
S17 & 0.618 & 0.874 & +0.257 \\
\hline
\textbf{Avg} & \textbf{0.646} & \textbf{0.685} & \textbf{+0.038} \\
\hline
\end{tabular}
\end{table}

\section{Analysis and Validation}

\subsection{Quantitative Analysis}

The most crucial observation from \ref{tab:parameter_summary} is the bimodal distribution of the optimal threshold percentiles among all the subjects. We can observe that the subjects diverged into two clusters based on their individual physiological reactivity to the stress stimuli. For subjects like S2, S3, S10, or S16 that showed a high physiological stress response, the grid search resulted in best-fitting percentiles in the 86th-93rd range. These cutoff points were feasible for the first cluster of subjects because these individuals had stable baseline resting heartbeats. For this reason, the grid search was able to find a clear dividing line that classified resting heartbeats from stress segments. 
On the other hand, subjects like S7, S8, S9, or S15 reverted back to the 50th percentile, because their cardiac morphology remained relatively stationary during both baseline and stress conditions. Due to the fact that there was a significant overlap between the two conditions, the grid search failed to find a distinct separating cutoff point between them. 
Failures like this weren’t caused by errors in the logic of the calibration process, but were rather a result of a real challenge posed by the unsupervised nature of the framework. Since the system assumes that the initial baseline recording includes a truly relaxed state of the subject, any previous physiological disruption can lead to tainted reference points. 
If we look at the metadata from the TSST experiment, it reveals that before the study, Subject 8 already experienced quite a stressful day and felt cold in the room where the protocol took place. Similarly, Subject 9 was feeling ill at the moment of the experiment, while Subject 7 participated in sports earlier that day, an activity which was specifically advised against by the WESAD researchers. These factors shed light on the non-responsiveness of the second cluster of subjects, as it becomes evident that their baseline heartbeats were already showing an aroused state. Because of this, the learned motifs didn’t truly reflect a resting cardiac state, causing the thresholds to default to 50.

Based on the results summarized in Table~\ref{tab:performance_results}, we can see how the subjects were categorized into three categories based on how strongly their cardiac morphology was subject to change during the stress condition. 
The first group of subjects such as S2, S3, S10, S14, or S16 achieved high F1-scores, which were above 0.85 with separation ratios of around 1.013.  For these specific subjects, their response to stress stimuli produced a prominent change in their cardiac morphology and timing. Furthermore, the increased heart rate caused multiple cardiac cycles to pack closer together, and the selected time window of 1.5 seconds caught multiple heartbeats, making the Matrix profile mark the segments as anomalies. Interestingly, the 3-second rolling average smoothing that was applied to the anomaly score proved to be efficient for this group as it removed any transient noise and helped the model preserve genuine stress responses. 
The second group of subjects, including S4, S5, or S6, displayed a moderate performance with an F1-score ranging between 0.60 and 0.79. For these particular subjects, the cardiac morphology shifted under the stress condition, but the change wasn’t as significant as in the first group of subjects. Since they showed a lower separation ratio, it indicates that the model had to distinguish between overlapping states of baseline and stress conditions. For this reason, the personalization of the percentile search proved to be effective, as otherwise the model would’ve struggled to identify the gap between the two states. 
The last group of subjects, like S8, S9, S11, or S15, includes the most challenging cases, where the F1-scores tend to be low, and the separation ratio is negligible. Here, Subject 8 represents the most extreme case, having a separation ratio of 0.99. This value indicates that for the specific subject, the model couldn’t signify any prevalent differences between relaxed and baseline cardiac states. As discussed previously, these failures can be attributed to a corrupted baseline signal for some subjects who were already stressed before the study started. 

Across all subjects, the model reached an overall average F1-score of 0.646 and an average accuracy of 0.668. These scores emphasize that each individual’s body has different physiological reactions to stress conditions. Moreover, we can observe that applying a common universal rule for threshold percentile selection (mean + 2 standard deviations of the baseline) produced an F1-score of only 0.242 and failed drastically for specific subjects like S4 or S5, which had a baseline variance that was higher than the common threshold. Together, these findings provide strong evidence that ultimately proves our hypothesis, which claims that physiological heart monitoring should be personalized to give meaningful results.

\subsection{Qualitative Analysis}

While the quantitative metrics comprehensively evaluate the framework’s performance, a qualitative visual analysis was also conducted to better illustrate how the Matrix Profile differentiates between the two states. By comparing the performance of a high responder like Subject 2 with the performance of a non-responder such as Subject 8, we can clearly identify what caused the framework to succeed in some cases and fail in others.

At first, we examine the baseline motifs extracted from the relaxed cardiac signal of the subjects, which can be seen in Fig.~\ref{fig:motifs_comparison}.

\begin{figure}[htbp]
    \centering
    \includegraphics[width=\columnwidth]{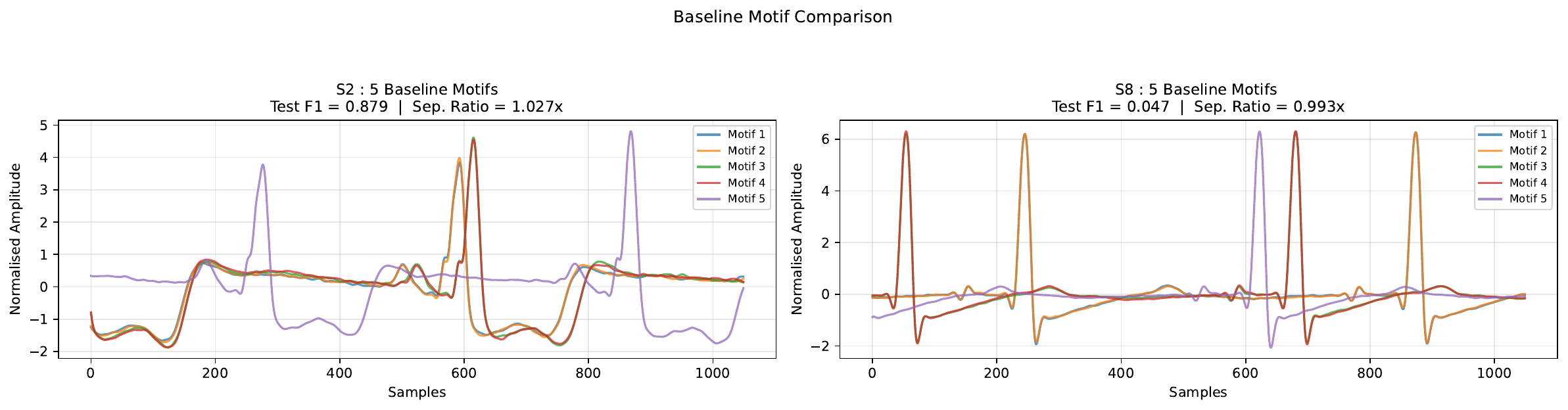}
    \caption{Baseline motif comparison between S2 (best performer, F1~=~0.879) and S8 (worst performer, F1~=~0.047).}
    \label{fig:motifs_comparison}
\end{figure}

Looking closely at Fig.~\ref{fig:motifs_comparison}, we can see that in the case of subject 2, the learned baseline motifs are nearly identical to one another, which can be seen in the way they overlap with each other with high consistency. There is some difference between them, but at the core, the shape is clear and aligned, indicating that the subject was truly relaxed during the baseline state. We can see the QRS complex as a massive, sharp peak, and the P and T waves before and after the massive spike. Therefore, the motifs, indeed, look like a healthy baseline signal. Thus, the Matrix Profile learned a highly reliable and accurate reference point for anomaly detection.

But, when we look at the learned motifs of Subject 8, it is evidently clear that the patterns are not only entirely different from one another, but also highly varied and disorganized. We see the presence of multiple QRS complexes within the span of 1.5 seconds, indicating an already aroused and racing heartbeat. As a result, the algorithm took an unstable signal as a reference and failed during execution. For this particular reason, we can conclude that the failure of Subject 2 can be mainly attributed to relying on a tainted signal as a baseline reference rather than to a flaw in the distance logic.

Following this, the second step of obtaining the qualitative results entails the examination of the exact moment the anomaly detection was maximized.

\begin{figure}[htbp]
    \centering
    \includegraphics[width=\columnwidth]{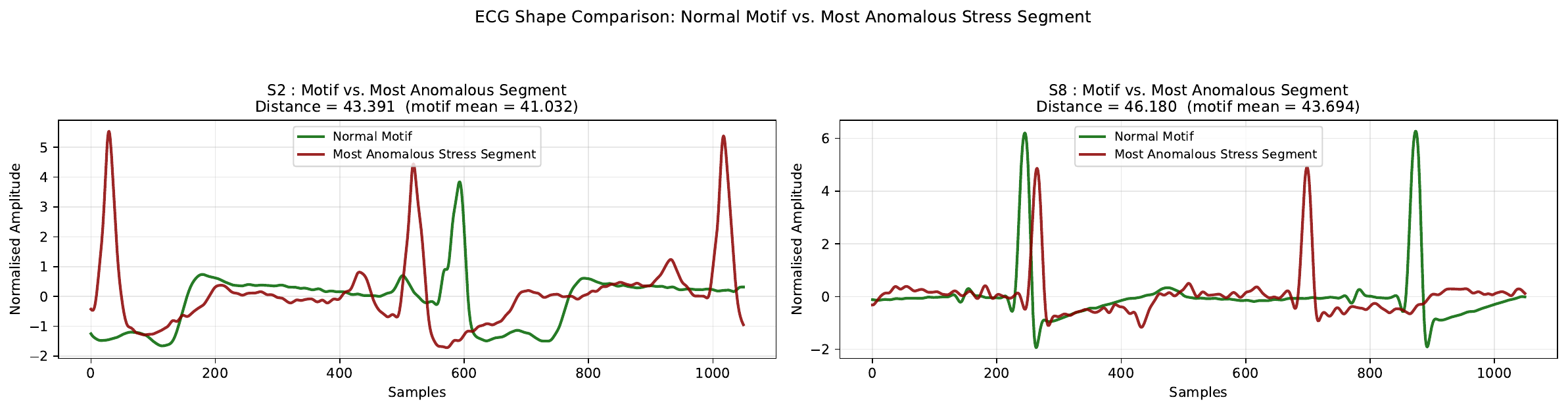}
    \caption{Baseline Motif vs Most Anomalous Segment Comparison between S2 (best performer, F1~=~0.879) and S8 (worst performer, F1~=~0.047).}
    \label{fig:shapes_comparison}
\end{figure}

As shown in Fig.~\ref{fig:shapes_comparison}, the baseline motif is being compared to the most anomalous segment captured during the stress protocol for both of the subjects.
The visual analysis of the structural shapes of Subject 2 shows a remarkable distinction between the baseline and stressed states. Specifically, we can observe how the most anomalous segments included almost two consecutive heartbeats within the time window of 1.5 seconds, indicating that the framework correctly flagged a stressed state as an anomaly and achieved high values during the evaluation stage. 
In comparison, looking at the structural shapes of Subject 8, it can be seen that the learned motif already contains two separate QRS complexes in a span of 1.5 seconds. Furthermore, looking at the most anomalous segment, we can see that the two cognitive states look nearly identical, thus hindering the framework’s ability to correctly differentiate anomalous segments from the baseline condition, resulting in an overall poor performance.

Fig.~\ref{fig:anomaly_detection_comparison} represents how the above-mentioned facts transform into distance tracking over time. 

In the case of Subject 2, we observe how, during the baseline state, the Matrix Profile distance remained consistently low. Momentarily, when the TSST experiment begins, after the 20-second transition point, the increased heart rate and morphological changes helped the anomaly score spike. Moreover, the alarm got triggered precisely when the change occurred, and the reason is that the personalized threshold was perfectly calibrated for the subject’s resting state. This percentile served as an optimal discriminator as the model ignored negligible fluctuations and became extremely sensitive to the packed heartbeats. 
In contrast, the visual dashboard of Subject 8 explains the exact reason for the detection failing for the particular subject. Here, when the stress protocol began, there was no substantial change in the cardiac behavior of that subject, and because of this, the anomaly score stayed entirely flat. After the 20-second transition, it couldn’t rise up, as a result of which, the threshold wasn’t crossed, and there was no alarm to be triggered.

\subsection{Supplementary Experiment Analysis}

As Table~\ref{tab:hrv_comparison} shows, there was no significant change in the average F1-score of the combined model. It changed moderately, going from 0.646 to 0.685. But when we look at the results of each subject separately, we can see that adding the HRV feature helped some subjects by improving their F1-score and weakened the results of other subjects. 
Both the high-performing and the moderate-performing subjects, such as S2, S3, S5, S11, etc., benefited from this experiment. For subjects like this, the RMSSD moved in the expected direction when the stress condition began, and, for that reason, it was easy for the model to correctly distinguish between their baseline and stress states. The Motif feature captured the changes in the shape of the heartbeat, while the HRV feature caught changes in the rhythm. 
Some previously poor-performing subjects, like S13, showed improvement under the supplementary experiment, which means that incorporating the HRV feature might be beneficial in certain cases.
For extremely poor-performing subjects, the F1-score suffered a substantial decline. Subjects like S9, S15, and S7 experienced an obvious drop in their metric values, and in the case of Subject 8, the F1-score even dropped to zero. This decrease can partially be attributed to the limitation discussed previously. Since for some subjects, their baseline recordings were corrupted, the model couldn’t distinguish between the two states despite the presence of another added feature. 
As a result, it can be concluded that adding other metrics like the RMSSD is not a common improvement and can only contribute to better results in certain individual cases. If some subjects have an already tainted baseline signal or some unusual individual physiological activity characteristics, the addition of a new feature can result in more noise and produce worse results.

\subsection{Robustness Check}

To make sure that the framework’s performance did not depend merely on a specific starting point in the data, we decided to shift the training, validation, and test windows forward by 30 seconds. This way, the algorithm was required to find and extract completely new baseline motifs and calculate a different optimal threshold for each subject. The purpose of this additional experiment was to check the robustness of the system and ensure that the previous results were not achieved as a result of accidental data alignment.
The new results demonstrated general consistency related to the two time windows. In the case of the shifted framework, an F1-score of 0.608 was achieved, which differed from the original pipeline by just 0.038 points. This subtle variance confirmed that the previous results were not captured by chance and were actual representations of the subjects’ stable cardiac behavior. For some high-performing and moderate-performing subjects like S3, S5, or S6, the difference between the two F1-scores was subtle and almost near zero. This means that their cardiac behavior stays consistent even with a different starting point in time.
Even though the pipeline was consistent for the majority of the subjects, in some cases, the shift revealed possible limitations of our primary approach. In the case of Subject 4, the performance of the framework dropped significantly by 0.55 points. Since this subject had a low separation ratio between the baseline and stress states, the method became highly sensitive to subtle changes or fluctuations that came with the time shift. So, this drop in performance was anticipated, and it reveals that for some subjects, acute stress doesn’t create drastic changes in the heartbeat morphology. This means that the efficiency of the approach is partially bounded by the subjects’ reactivity to the stress-producing stimuli.
The shifted pipeline also showed that for some subjects whose baseline data might have been corrupted by noise, the shift can become beneficial. For example, Subject 13 showed an improvement of 0.09 points. This can be attributed to the fact that the first time window may have captured noise or arousal before the TSST experiment, which could have caused low separation between stress and baseline states. The shift in time allowed the framework to learn from cleaner baseline references, thus resulting in improved performance.

\subsection{Limitations of the work}

Even though our suggested framework addresses the gap of interpretability, it has several limitations that need to be taken into consideration.
First, the efficient performance of the method depends heavily on the assumption that stress produces noticeable changes in heartbeat shape that are different from baseline conditions. While we proved that this assumption is true, it may not hold for subtle or gradual stress, which develops over a long time and doesn’t show a sharp physiological shift.
Another limitation relates to the hyperparameter sensitivity of the framework. As it was stated in previous sections, motif discovery can be quite sensitive to choices of window sizes and there is no universally accepted approach for choosing the optimal one. Whilst a window size of 1.5 seconds allows the system to capture a full cardiac waveform for the majority, it may still struggle to capture some gradual changes, for which a longer time frame is needed.
Lastly, even though the framework helps with interpretability and is accessible for human understanding, it still can’t provide exact clinical diagnoses, and the opinion of healthcare experts is crucial. The framework can offer the perspective that the cardiac waveform has deviated from normal behavior, but it can’t differentiate between stress, arrhythmia, or other conditions. 

\section{Discussion and Conclusion}

\subsection{Main Contributions}

The core objective of this research was to assess whether a completely unsupervised method could efficiently predict stress from recorded ECG signals in such a way that is both personalized and interpretable. The suggested approach addresses this challenge and demonstrates that it is feasible to detect changes in cardiac behavior as anomalies without relying on “black-box” models.
The main contribution of this work is the successful implementation of a stress detection pipeline, which learns what normal cardiac activity looks like for each subject specifically and then identifies deviations from that normal activity as potential indicators of stress. Yet, the most significant takeaway is that the framework requires no recordings of labeled stressed data during the training phase of the implementation. The only thing the pipeline needs is a clean, stable baseline ECG data. Due to this fact, the system can be utilized in real-life practical scenarios where stress should be continuously monitored, and it’s impossible to have predefined labels. 
Another contribution of the research lies in the evaluation method, which was conducted with the help of a strict procedure that split the data into three parts: training, validation, and testing sets. In the majority of studies conducted on the WESAD dataset, a cross-validation method was used. That helped the models use patterns from the whole group of subjects during the learning phase. In comparison, our framework has no overlap between the phases. This means that the threshold that was calibrated during the validation phase didn’t have any role in the baseline motif learning process. Consequently, the testing data remained fully unseen before the final evaluation process began. This way, our model becomes adaptable in real-time settings when it must adjust to a new subject whose stress responses weren’t prerecorded and are unknown.
Finally, when we examine the cases where the method failed and where it succeeded, this study provides a comprehensive explanation of when exactly this kind of detection can be trusted. Perhaps, the most interesting contribution and discovery of the work is that the way a person’s body reacts to the stress and how clean their initial baseline data is, matters more than how complex the model’s architecture is or how much hyperparameter tuning has been done.

\subsection{Key Findings}

Besides demonstrating an unsupervised motif-based stress detection system, the results of our study offer various key findings and unique advantages that come with personalization and interpretability. 
The first and foremost finding is that personalization is not an optional strategy, but rather a necessary approach in such problems. The quantitative analysis of the study proves that using a personalized threshold calibration for each subject resulted in a major performance improvement compared to the universal common approach. This essentially proves the most crucial hypothesis of the research. The deciding factor in stress detection accuracy is individual physiological variability, and not the complexity of the model. 
Another key finding was that the performance and the accuracy of detection were also dependent on the quality and the integrity of the baseline stable data. In the Results section, it was revealed that subjects who experienced stress before the experiment had a corrupted baseline signal, which prevented the model from accurately learning between baseline and stress signals. In contrast to this, subjects who had clean and healthy baseline data resulted in a more efficient performance of the model, achieving F1-scores of up to 0.89. 

The research also achieved another one of its important objectives, which was providing easily interpretable signal-level explanations of stress detection. All of the decisions made by the model can be visually explained by simply overlaying the ECG segment flagged as stress on top of the learned baseline motif and comparing their shapes. The motifs closely resemble the typical characteristics of an ECG signal, such as the P-QRS-T waveforms. Any difference from this can be recognized as stress and can be easily comprehensible to the human eye. Compared to other supervised machine learning or deep learning models, this approach transforms a mathematical score into a visually understandable story. 

Lastly, according to our study, it was revealed that the addition of an additional feature like the RMSSD doesn’t always guarantee more accurate or efficient results. For the subjects who had more stable heartbeats, the new feature provided an increase in model performance. However, that wasn’t the case with subjects whose baseline data was already corrupted. This suggests that the inclusion of other features should be applied selectively and after a high-quality, clear baseline data is established.

\subsection{Future Improvements}

The most essential next step for this framework is to add an automatic baseline quality verification step that will run before the system starts. The model’s efficient performance relied almost entirely on having genuinely clean and calm baseline data. For that reason, this verification step would measure how consistent the learned motifs are and, in case of being unstable, it would reject the baseline and ask the user to record another one. 
Another way to help the initially corrupted baseline would be the implementation of a dynamic baseline update mechanism. This way, the system could use the recently learned ECG patterns and periodically update its reference motifs. In this way, the model could become more robust without sacrificing the personalization approach, which is what makes the approach effective.
An additional direction to explore would be to implement a carefully chosen smart combination of signals. Instead of adding additional features like RMSSD for everyone, the system could decide whether to utilize it depending on the quality of the baseline data and the individual responses of subjects. In this way, the performance of the framework could be improved in case of non-responders without overriding the interpretability of the Matrix Profile.

\subsection{Conclusion}

This research aimed to achieve three main objectives, which were interlinked with one another. We developed a fully unsupervised, interpretable system for stress detection from ECG signals using a motif-based anomaly analysis, which was also personalized for each subject. By using the Matrix Profile approach, the framework learned cardiac patterns uniquely for each person from unlabeled baseline data and detected stress as a morphological divergence from those reference patterns. Each detection was then shown visually and justified through the comparison of ECG waveforms, which allowed it to be transparent, reliable, and easily understandable. 
The efficiency of the method was validated through splitting data into train, validation, and test splits, which ensured there was no overfitting to calibration data and the performance had real generalization.
The final results of the research confirm that the core objectives were achieved successfully for subjects with reliable and healthy baseline recordings, where F1-scores of 0.85 and higher were achieved. Furthermore, the results also validate the advantage of the personalized approach over a generalized one, resulting in much higher accuracy and F1-score in the first case.
The framework struggled to efficiently operate for specific subjects who had a corrupted initial baseline recording, as they were stressed before the TSST protocol even began.
These findings, combined, offer an honest perspective on the trade-off between personalization and accuracy. That perspective establishes that a motif-based anomaly detection can be a more transparent alternative for physiological monitoring of stress.

\section*{Conflict of Interest} The authors declare no conflict of interest.

\end{document}